\documentclass[10pt,conf]{IEEEtran}
\usepackage{geometry}
\usepackage{bbold}
\usepackage{float}
\usepackage{subcaption}%
\usepackage[cmex10]{amsmath}
\usepackage{amssymb}
\usepackage{verbatim}
\usepackage{array}
\usepackage{latexsym}
\usepackage{color}
\usepackage{tabularx} 
\usepackage{textgreek}
\usepackage{subscript}
\usepackage{rotating}
\usepackage{booktabs,multirow}
\usepackage{mdframed}	
\usepackage{tabularx}
\usepackage{amsmath}
\usepackage{makecell}
\usepackage{tabto}

\usepackage{mathrsfs}
\usepackage{url}
\usepackage{cite}
\usepackage{listings}
\usepackage{array}
\usepackage[table]{xcolor}

\newcommand{\nref}[1]{\textcolor{red}{[{\bf NEED CITATION}]}}
\begin{document}
\setlength{\columnsep}{0.2 in}
\def\BibTeX{{\rm B\kern-.05em{\sc i\kern-.025em b}\kern-.08em T\kern-.1667em\lower.7ex\hbox{E}\kern-.125emX}}

\title{A Reliable and Efficient 5G Vehicular MEC: Guaranteed Task Completion with Minimal Latency}

\author{
	Mahsa~Paknejad,~Parisa~Fard~Moshiri,~Murat~Simsek~\IEEEmembership{Senior~Member,~IEEE},\\~Burak Kantarci,~\IEEEmembership{Senior Member,~IEEE}~and~Hussein~T.~Mouftah,~\IEEEmembership{Fellow,~IEEE}
\thanks{
The authors are with the School of Electrical Engineering and Computer Science at the University of Ottawa, Ottawa, ON, K1N 6N5, Canada.
E-mail: \{mahsa.paknejad, parisa.fard.moshiri, murat.simsek, burak.kantarci, mouftah\}@uottawa.ca}  
}

\maketitle
\thispagestyle{empty}
\pagestyle{empty}
\begin{abstract}
This paper explores the advancement of Vehicular Edge Computing (VEC) as a tailored application of Mobile Edge Computing (MEC) for the automotive industry, addressing the rising demand for real-time processing in connected and autonomous vehicles. VEC brings computational resources closer to vehicles, reducing data processing delays crucial for safety-critical applications such as autonomous driving and intelligent traffic management. However, the challenge lies in managing the high and dynamic task load generated by vehicles’ data streams. We focus on enhancing task offloading and scheduling techniques to optimize both communication and computation latencies in VEC networks. Our approach involves implementing task scheduling algorithms, including First-Come, First-Served (FCFS), Shortest Deadline First (SDF), and Particle Swarm Optimization (PSO) for optimization. Additionally, we divide portions of tasks between the MEC servers and vehicles to reduce the number of dropped tasks and improve real-time adaptability. This paper also compares fixed and shared bandwidth scenarios to manage transmission efficiency under varying loads. Our findings indicate that MEC+Local (partitioning) scenario significantly outperforms MEC-only scenario by ensuring the completion of all tasks, resulting in a zero task drop ratio. The MEC-only scenario demonstrates approximately 5.65\% better average end-to-end latency compared to the MEC+Local (partitioning) scenario when handling 200 tasks. However, this improvement comes at the cost of dropping a significant number of tasks (109 out of 200). Additionally, allocating shared bandwidth helps to slightly decrease transmission waiting time compared to using fixed bandwidth.
\end{abstract}
\begin{IEEEkeywords}
Mobile edge computing, task offloading, 5G, optimization. 
\end{IEEEkeywords}

\IEEEpeerreviewmaketitle

\section{Introduction}
Mobile Edge Computing (MEC) has gained increasing importance in today's technology landscape, where the number of connected devices continues to grow. By bringing computational power closer to where data is generated, MEC minimizes the need to send information to distant cloud servers, reducing data processing delays \cite{ferrag.comst.2023}. 
Building on MEC's benefits, Vehicular Edge Computing (VEC) tailors this concept specifically for the automotive and transportation industries. VEC brings processing capabilities closer to vehicles, enabling faster handling of data from sensors and communication systems in connected vehicles \cite{vec2023}. This edge-based processing is crucial for applications like autonomous driving, where real-time decision-making is vital for safety \cite{autodriving2019}. By providing essential computing resources at the network's edge, VEC ensures that connected vehicles operate more safely, efficiently, and reliably within today’s transportation networks \cite{meson2024}. Vehicles continuously send data that require fast processing, which can overwhelm edge servers and cause communication and processing delays \cite{AVCIL2024100773}
Inadequate task management may result in task drops by servers; however, optimizing the task queue can alleviate these delays by establishing the most efficient processing sequence.

MEC+Local (partitioning) approach  also improves the efficiency of the VEC by dividing tasks between the MEC servers and the vehicles themselves \cite{partition2023}. This approach leverages computing resources at both the edge and within vehicles to reduce processing time and improve flexibility. This balanced approach prevents task drops and ensures that VEC systems handle the workload effectively. The goal is to offload as many tasks as possible to the VEC without overwhelming it, which could result in some tasks not completing on time. By balancing local and edge processing, MEC+Local (partitioning) approach  reduces communication and computation delays, ensuring tasks are completed on time. This approach enhances system efficiency, supporting the advancement of sophisticated VEC applications designed to meet the needs of modern transportation \cite{Cloud-Edge2021} \cite{online2023}.
The main contributions of this paper are as follows:
\begin{enumerate}
\item  Our scenario includes strict deadlines based on Roadside Unit (RSU) coverage and large tasks, offering a more accurate assessment of network performance under high-stress conditions.

\item  Simultaneous tasks can share channels for uplink and downlink, thereby reducing transmission waiting times.

\item  By tracking and analyzing dropped tasks within the RSU, the system dynamically adjusts parameters to reduce task loss and optimize resource use.

\end{enumerate}

Section II presents the literature review, Section III the system model, Section IV the performance analysis, and Section V the conclusions.

\section{Related Work}
Recent studies have introduced a variety of strategies to improve task offloading in MEC for vehicular networks.
Li et al. \cite{relate1} investigate task offloading using deep reinforcement learning (DRL) within MEC-supported heterogeneous vehicular networks, utilizing both V2I and V2V communications. They aim to maximize computation rates using serial and parallel offloading schemes. However, their study mainly focuses on computation optimization, with limited emphasis on communication resource management, potentially causing bottlenecks in high-load environments. Moreover, their model lacks scalable MEC+Local (partitioning) approach and effective dropped task management, which are crucial for high-demand, time-sensitive applications. Similarly, Li and Fan \cite{relate2} propose a NOMA-assisted MEC model that emphasizes computation resource management in mobility-aware networks. By considering vehicle speeds and task arrival rates, they work to optimize task offloading performance. Although they employ NOMA for channel sharing, their model’s scalability is limited, assuming a fixed setup where each subchannel is shared by only two vehicles. Additionally, the model does not address task partitioning or advanced communication strategies that could adapt to varying traffic conditions. Another study by Li et al. \cite{relate6} presents a DRL-based collaborative framework for vehicular networks, using DDPG techniques to improve task offloading and reduce service delays by distributing tasks in parallel. Despite this model’s adaptability to dynamic vehicular environments, it still lacks flexible communication resource allocation and options for local processing, which could enhance its adaptability and decrease delays in constrained network scenarios. Building on previous work, Moon and Lim \cite{relate5} propose a framework that focuses on task partitioning and migration across MEC servers to improve load balancing. Their model effectively reduces task delays by migrating tasks from overloaded servers to those with lighter loads; however, it still lacks a flexible approach to communication resource management. Similar to \cite{relate6}, it does not consider the potential for local processing within vehicles, which could provide added flexibility when MEC resources are under strain. Further research on MEC optimization \cite{relate3} explores joint computation offloading and resource allocation, incorporating partial and MEC-only offloading schemes to reduce latency. While task partitioning proves beneficial for latency reduction, the model’s dependence on OFDMA restricts communication flexibility and does not include management of the number of dropped tasks. Expanding upon previous studies, Dai et al. \cite{relate4} present a joint optimization framework that addresses offloading, resource allocation, and data caching with the goal of minimizing processing costs. Utilizing a Binary Particle Swarm Optimization (PSO) algorithm, they target tasks that need both computation and data caching. Nonetheless, the model does not include dynamic channel sharing or number of dropped tasks considerations.

To address these existing limitations, we aim to develop a more realistic simulation that models scenarios with strict deadlines and large task sizes. In addition to optimizing computation, we enhance communication processes for both uplink and downlink, considering the impact of dropped tasks to enhance the overall performance. For the communication aspect, we manage bandwidth sharing by enabling tasks that arrive simultaneously within the RSU range to share bandwidth during uplink, as well as for tasks that are ready for parallel downlink. This approach aims to reduce transmission waiting time, thereby improving efficiency across the network.

\section{System Model}
\subsection{The Offloading Schemes}
We explore scheduling algorithms including First-Come, First-Served (FCFS), Shortest Deadline First (SDF), and a heuristic approach, PSO. FCFS processes tasks strictly by arrival order, offering simplicity and predictability but ignoring task deadlines \cite{parisa}. This can delay time-sensitive tasks and lead to inefficiencies, increased latency, and task drops, especially under heavy traffic. The SDF scheduling method prioritizes tasks based on deadlines, with shorter deadlines processed first. Deadlines depend on the distance between vehicles and the RSU, allowing tasks with less remaining time to be prioritized and reducing dropped tasks. However, SDF can delay tasks with longer deadlines, especially in high-traffic situations. 

\begin{figure*}[h]
    \centering
    \includegraphics[width=0.7\textwidth, clip, trim= 0.0cm 0.0cm 0.5cm 1.2cm]{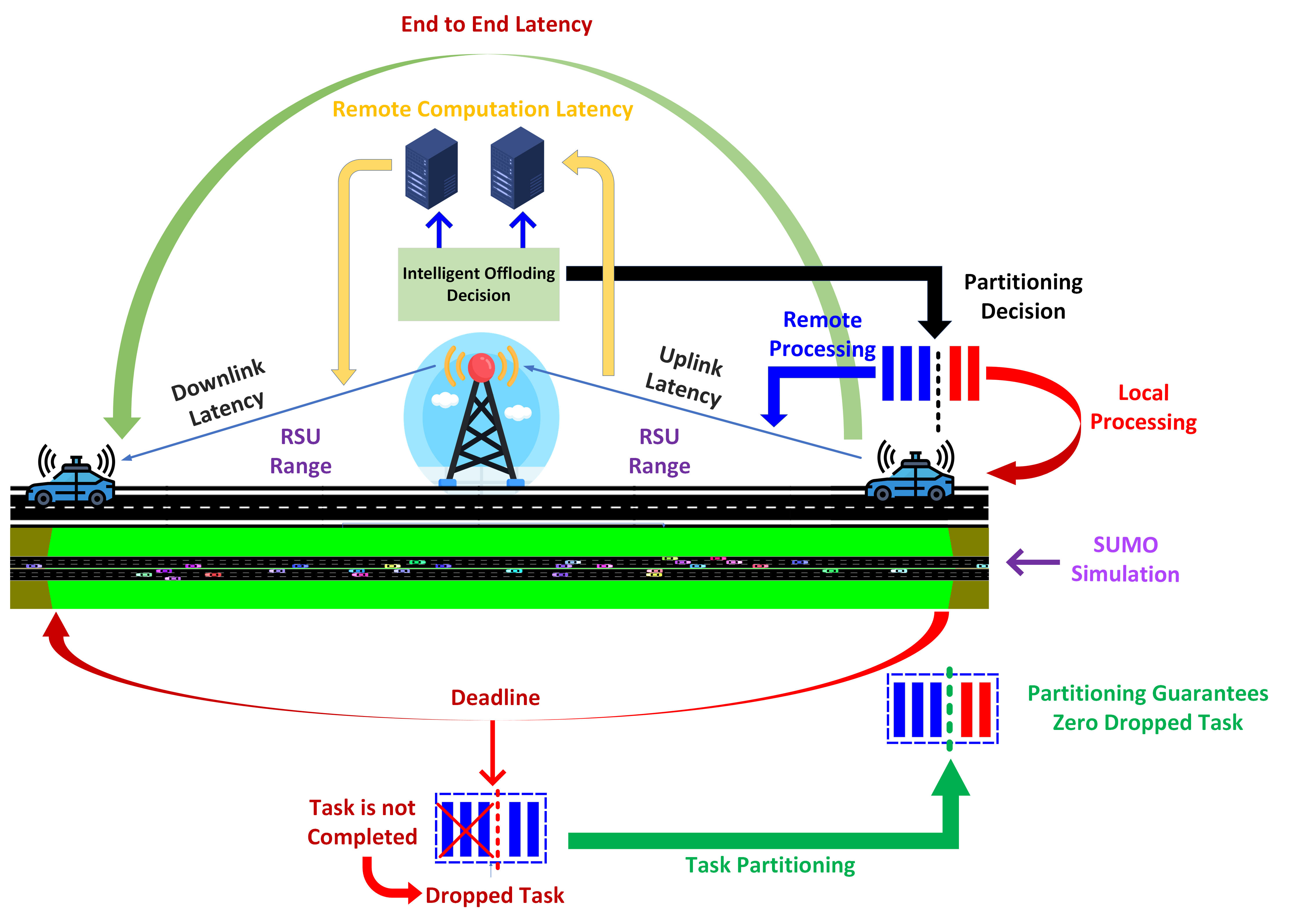}
    \caption{Problem statement for MEC+Local (partitioning) and MEC-only scenarios in 5G Vehicular MEC}
    \label{fig:scenario}
\end{figure*}

\begin{table}[!hbt]
\centering
\renewcommand{\arraystretch}{1.2} 
 
\caption{Notation table}
\label{tab:notation}
\begin{tabularx}
{0.43\textwidth}{|c|X|}
\hline
\textbf{Parameter} & \textbf{Description} \\
\hline
\( L_i^{\text{cm}} \) & Communication latency for task \(i\)\\
\hline
\( T_i^{\text{cm}} \) & Communication time for task \(i\)\\
\hline
\( W_i^{\text{u}} \) & uplink waiting time for task \(i\)\\
\hline
\( W_i^{\text{d}} \) & Downlink Waiting time for task \(i\)\\
\hline
\( S_i \) & Size of task \( i \)\\
\hline
\( r_i \) & transmission rate for task \( i \)\\
\hline
\( B_i \) & Bandwidth of task \( i \) \\
\hline
\( p \) & Transmission power \\
\hline
\( g \) & Channel gain \\
\hline
\( n_0 \) & Density of noise power\\
\hline
\(t_{i}^{\text{a}}\) & Arrival time of task \( i \) on the RSU\\
\hline
\(t_{i}^{\text{a}'}\) & Arrival time of task \( i \) to the vehicle \\
\hline
\(t_{i}^{\text{r}}\) & The time at which task \(i\) is ready to be offloaded to the RSU\\
\hline
\(t_{i}^{\text{r}'}\) & The time at which task \(i\) is finished processing and ready to be sent back to its vehicle\\
\hline
\(t_{i}^{\text{sp}}\) & Start processing time for task \( i \) \\
\hline
\( B_{\text{max}}\) & Maximum bandwidth \\
\hline
\(L_{i}^{\text{p}} \) & Remote computation latency for task \( i \)\\
\hline
$T_{i}^{\text{p}}$ & Remote computation time for task \( i \)\\
\hline
\(T_{i}^{\text{pl}}\) & Local computation time for task \( i \) \\
\hline
\(W_{i}^{\text{p}}\) & Remote computation waiting time for task \( i \)\\
\hline

$n$ & Total number of tasks\\
\hline
$n'$ &  Number of tasks that arrive at the same time at RSU range or finish processing on each MEC server simultaneously\\
\hline
$m$ & Total number of MEC servers\\
\hline
$x_{ij}$ & Binary decision variable for assigning task \(i\) to MEC \(j\)\\
\hline

\(T_{i}^{\text{range}}\) & Remaining time in the RSU range for task \( i \)\\
\hline
 
\end{tabularx}
\label{tab:notations} 
\end{table} 

Our optimization approach includes both MEC-only and MEC+Local (partitioning) approaches. In the MEC-only method, the entire task is offloaded to a MEC server, whereas MEC+Local (partitioning) approach divides the task such that some components are processed remotely while others are handled locally on the vehicle.  This approach is beneficial when a significant number of tasks are dropped in the MEC-only scenario, as it ensures that no tasks are dropped. Based on \figurename~\ref{fig:scenario}, the offloading scenario is simulated using Simulation of Urban MObility (SUMO). In this simulation, vehicles offload computationally intensive tasks to the RSU when they enter its communication range. Task deadlines are defined by the duration that vehicles remain within the RSU's coverage. Based on decisions made by the PSO algorithm, tasks are partitioned, with some portions offloaded to the RSU and the remainder processed locally. PSO determines the optimal queue position for each task and assigns it to the most suitable MEC server. The algorithm focuses on minimizing overall latency and reducing the number of dropped tasks by evaluating factors such as task deadlines and server availability. Based on PSO decisions, portions of tasks that cannot be processed and returned to the vehicle before it leaves the RSU’s range are either processed locally in MEC+Local (partitioning) approach or dropped in MEC-only scenario. This decision-making process takes place at the RSU.

The communication latency includes both the time taken to send tasks from the vehicles to the RSU, and the time taken for the vehicles to receive results from the RSU.
Moreover, we consider two scenarios for bandwidth allocation. In the first scenario, all tasks have a fixed maximum bandwidth, so only one task is offloaded at any given time. In the second scenario, when multiple vehicles arrive within the RSU range with tasks to be offloaded, those tasks are offloaded simultaneously by sharing the available bandwidth. These approaches are also considered in the downlink phase, where results are transmitted back to vehicles from the MEC servers.


\subsection{Optimization}
All mathematical notations are provided in Table~\ref{tab:notations}.
As illustrated in Figure~\ref{fig:scenario}, the end-to-end latency is the sum of communication latency, computation latency, and local computation. The total communication latency is formulated in \eqref{eq:Lcm}. Since the transmission time, \(T_i^{\text{cm}}\), is identical for both uplink and downlink, given that the task output sizes are assumed to match their input sizes, \(T_i^{\text{cm}}\) is multiplied by 2 to account for both directions. \(T_i^{\text{cm}} \) is formulated in \eqref{eq:Tcm}, with \( r_i \) defined in \eqref{eq:r}.

\begin{equation}
L_i^{\text{cm}} = 2 \times T_i^{\text{cm}} + W_i^{\text{u}} + W_{i}^{\text{d}} 
    \label{eq:Lcm}
\end{equation}




\begin{equation}
T_i^{\text{cm}}  = \frac{S_i}{r_i}
    \label{eq:Tcm}
\end{equation}

\begin{equation}
r_i = B_i \cdot \log_2 \left( 1 + \frac{p \cdot g}{n_0} \right)
    \label{eq:r}
\end{equation} 

\(W_i^{\text{u}}\) is formulated in \eqref{eq:Wu}, which represents the time each task spends waiting in a vehicle until bandwidth is available.

\begin{equation}
W_i^{\text{u}} = t_{i-1}^{\text{a}} - t_{i}^{\text{r}}
    \label{eq:Wu}
\end{equation}

In scenarios where a vehicle enters the coverage area while a prior task is still being transmitted to the RSU, it must wait until the bandwidth is totally available. However, if a task is the first to be offloaded or is ready after the previous task has fully arrived, \( W_i^{\text{u}} \) is zero.

Furthermore, we consider  \(W_i^{\text{d}}\) for each MEC server when transmitting results back to the vehicles. As formulated in \eqref{eq:Wd}, if a task completes processing before the prior task has been fully delivered to the vehicle, it must wait until the bandwidth is available.

\begin{equation}
W_{i}^{\text{d}} = t_{i-1}^{\text{a}'} - t_{i}^{\text{r}'}
    \label{eq:Wd}
\end{equation} 

Similar to \(W_i^{\text{u}}\), if a task is the first one or becomes ready to be sent back to the vehicle after the previous task has fully arrived at its vehicle, \( W_{i}^{\text{d}} \) is set to zero. We also consider the potential waiting time in scenarios where tasks complete simultaneously on both servers. In such cases, one of the MEC servers experiences an additional delay of \( W_{i}^{\text{d}} \), determined by the task on the other MEC server. In the fixed bandwidth scenario, a maximum bandwidth is allocated for all tasks during both uplink and downlink. In contrast, the shared bandwidth approach allows bandwidth to be distributed among the tasks. According to \eqref{eq:B}, for both uplink and downlink, tasks that enter the RSU range simultaneously, as well as those completed at the same time on both MEC servers, will share the available bandwidth and be transmitted concurrently. Whereas, other tasks utilize the maximum bandwidth.

\begin{equation}
B_{i}^{\text{}} = 
\begin{cases} 
B_{\text{max}} \cdot \frac{S_{i}^{\text{}}}{n'}& \text{for } {n'}  \\ 
B_{\text{max}}  & \text{otherwise}
\end{cases}
    \label{eq:B}
\end{equation} 
In  \eqref{eq:B}, \(n'\) represents the number of tasks that either arrive simultaneously within the RSU range or are completed simultaneously on the MEC servers. Since there are only two MEC servers, \(n'=2\) for the downlink process. It is notable that tasks in each MEC server are completed sequentially, meaning that, in both scenarios, only one task from each MEC server utilizes the bandwidth at any given time. The computation latency for remote processing is represented as:

\begin{equation}
L_{i}^{\text{p}} = T_{i}^{\text{p}} + W_{i}^{\text{p}}
    \label{eq:Lcp}
\end{equation} 
\begin{equation}
W_{i}^{\text{p}} = t_{i}^{\text{sp}} - t_{i}^{\text{a}}
    \label{eq:Wcp}
\end{equation}

\( T_{i}^{\text{p}} \) and \( T_{i}^{\text{pl}} \) for each task are determined based on remote and local inference times from \cite{inference}. The formula for computation waiting time, \( W_{i}^{\text{p}} \) is provided in \eqref{eq:Wcp}.

  The start processing time of the first and second tasks, \( t_{1}^{\text{sp}} \), aligns with their arrival time at the RSU. This is because they are assigned to MEC servers as soon as they arrive. The \( t_{i}^{\text{sp}} \) is calculated as: 

\begin{equation}
t_{i}^{\text{sp}} = t_{i-1}^{\text{sp}} + T_{i-1}^{\text{p}}  
    \label{eq:sp}
\end{equation} Where for subsequent tasks after the first and second tasks, their start processing time, \( t_{i}^{\text{sp}} \) is the sum of their previous task's start processing time, \( t_{i-1}^{\text{sp}} \) and processing time, \(T_{i-1}^{\text{cp}}\). Our objective is to reduce both the total end-to-end latency and the number of dropped tasks, as illustrated in \eqref{eq:obj}. 
Our main focus is to first reduce the number of dropped tasks and then minimize the end-to-end latency of the remaining ones.

\begin{equation}
\text{min} \left( \lambda({\sum_{j=1}^{m} \sum_{i=1}^{n_{\text{}}} L_{i}^{\text{e2e}} \times x_{ij}) +  (1 - \lambda) \text{D}}\right)
    \label{eq:obj}
\end{equation} 
Where \(\lambda\) is the weight for end-to-end latency, while \(1 - \lambda\) represents the weight for the number of dropped tasks. \( L_{i}^{\text{e2e}} \) and \( \text{D} \) are defined as:

\begin{equation}
L_{i}^{\text{e2e}} = p_{i} \cdot (L_{i}^{\text{cm}} + L_{i}^{\text{p}}) + (1 - p_{i}) \cdot T_{i}^{\text{pl}}
    \label{eq:end2end}
\end{equation}

\begin{equation}
\text{D} ={\sum_{j=1}^{m} \sum_{i=1}^{n} (1 - x_{ij})}
    \label{eq:drop}
\end{equation} \(x_{ij}\) is indicated as:

\begin{equation}
x_{ij} = 
\begin{cases} 
1 & \text{if task } i \text{ is assigned to MEC } j \\ 
0 & \text{otherwise} 
\end{cases}
    \label{eq:assign}
\end{equation}

Also, \( p_i \) represents the fraction of a task that is offloaded to a MEC server, while \( (1 - p_i) \) denotes the portions processed locally. In MEC-only offloading, \( p_i \) is set to 1, indicating that the entire task is offloaded with no local processing. Our PSO algorithm designates \( p_i \) along with task indices, specifying their processing order. For MEC+Local (partitioning) scenario, we have \(x_{ij}\)=1 and $0 \leq p_i \leq 1$.
Additionally, the PSO determines the allocation of each task to a specific MEC server. According to constraint~\eqref{eq:assign-con}, each task is processed by only one MEC. Furthermore, constraint~\eqref{eq:inequality} ensures a task is only assigned to a server if the expected end-to-end latency is less or equal to remaining time in the range. 
\begin{equation}
\sum_{j=1}^{m} x_{ij} \leq 1 \quad \forall i \in {n_{}^{\text{}}}
    \label{eq:assign-con}
\end{equation}

\begin{equation}
1 - \left( \frac{L_{i}^{\text{e2e}}}{T_{i}^{\text{range}}} \right) \leq x_{ij}
\label{eq:inequality}
\end{equation}
This helps in managing the quality of service (QoS) by ensuring that the end-to-end latency does not exceed the remaining time in range for each task.
\section{Performance Analysis}

In this study, we simulate a highway scenario using the SUMO platform, where a RSU supports computational offloading for vehicles traveling along the highway. The RSU is equipped with two MEC servers, designed to handle incoming tasks from vehicles in its coverage area. We incorporate a time-sensitive requirement for each task, with deadlines determined by the Euclidean distance between the vehicle and the RSU at the time of task offloading. The simulation covers three different scenarios with varying traffic densities to evaluate MEC server performance under different workloads. Each vehicle generates a single image-processing task. Task generation follows a Poisson distribution to capture the randomness and variability of real-world vehicular task offloading. Tasks can be generated both when a vehicle is within the RSU’s range and any time prior to entering this range, which enables a more realistic model of real-world scenarios. The size of each task is determined by the resolution of the image being offloaded. We simulate different image resolutions to represent varying levels of computational load.

\begin{table}[!hbt]
\centering
\renewcommand{\arraystretch}{1} 
\caption{Parameters values in simulations} 
\begin{tabular}{|c|c|}
\hline
\textbf{Parameters} & \textbf{Value}   \\   \hline
Number of vehicles & 50, 100, 200   \\   \hline
Number of tasks per vehicle & 1 \\   \hline
Max bandwidth & 20Mhz   \\   \hline
Noise power & 100dbm   \\   \hline
Transmit power & 200 mW \\   \hline
Number of MEC servers & 2   \\   \hline
Number of CPU in each server & 1   \\   \hline
$\lambda$ & 0.3 \\
\hline
Swarm Size & 50 \\
\hline
Maximum number of iterations & 100 \\
\hline
Personal \& global Learning Coefficient & 0.5 \\
\hline
\end{tabular}
\label{tab:parameters} 
\end{table}

\begin{figure}[!hbt]
    \centering
    \begin{subfigure}[!hbt]{0.35\textwidth}
        \centering
        \includegraphics[width=\textwidth, trim=0.0cm 0.0cm 2cm 0.0cm, clip]{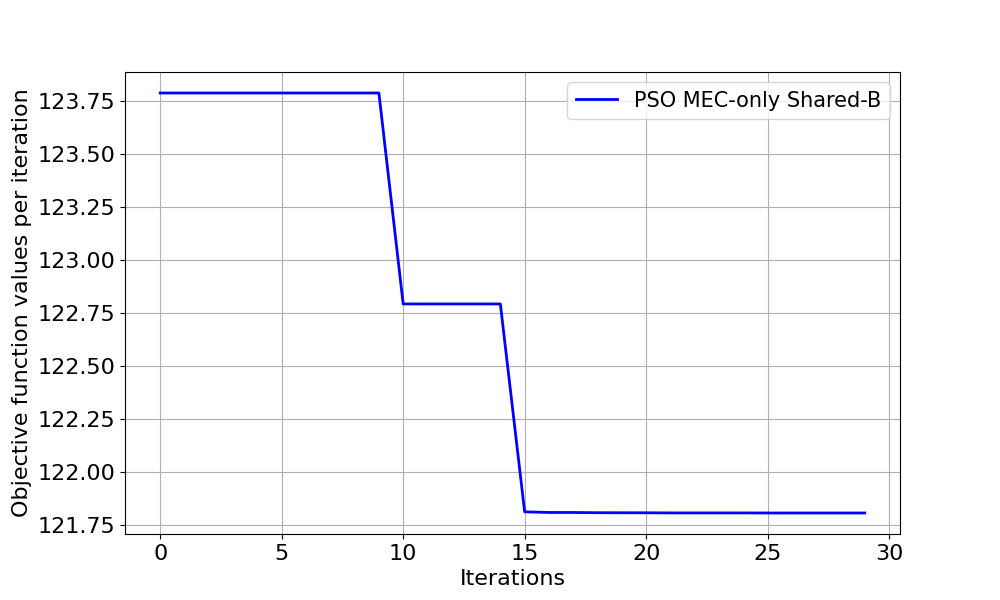}
        \caption{MEC-only Scenario}
        \label{fig:Con_mec_only}
    \end{subfigure}
    \hfill
    \begin{subfigure}[!hbt]{0.35\textwidth}
        \centering
        \includegraphics[width=\textwidth, trim=0.0cm 0.0cm 2cm 0.0cm, clip]{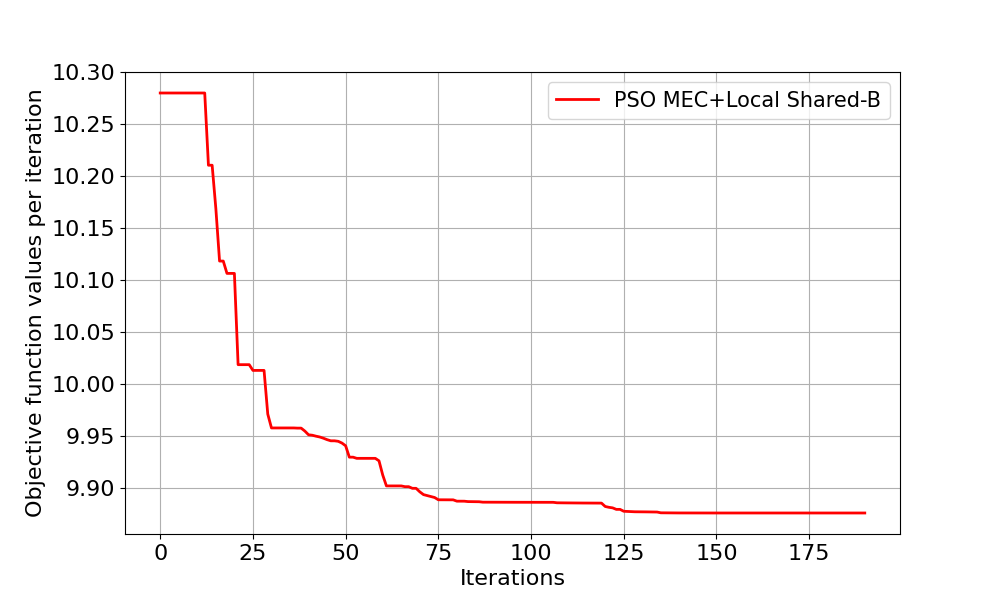}
        \caption{MEC+Local (task partitioning) Scenario}
        \label{fig:Con_partial}
    \end{subfigure}
    \caption{Convergence Plots for PSO with Shared Bandwidth for 200 tasks}
    \label{fig:Convergence_Plots}
\end{figure}

\begin{figure}[!hbt]
    \centering
    \includegraphics[width=0.35\textwidth, trim=1.2cm 0.0cm 2cm 1.2cm, clip]{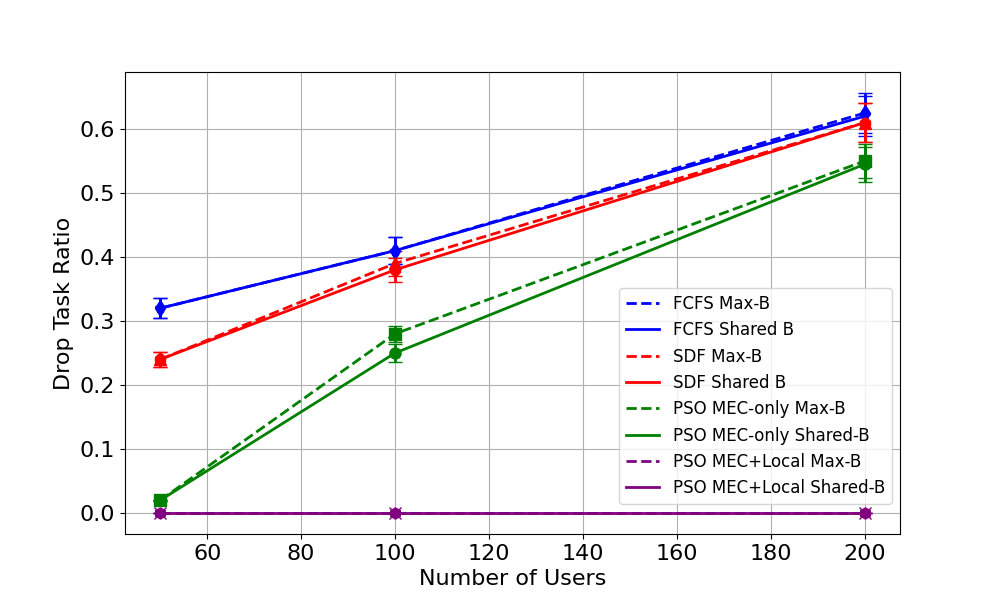}
    \caption{Average Drop task ratio for different number of users }
    \label{fig:Drop Task Ratio}
\end{figure}

\begin{figure}[!hbt]
    \centering
    \includegraphics[width=0.35\textwidth, trim=1.2cm 0.0cm 2cm 1.2cm, clip]{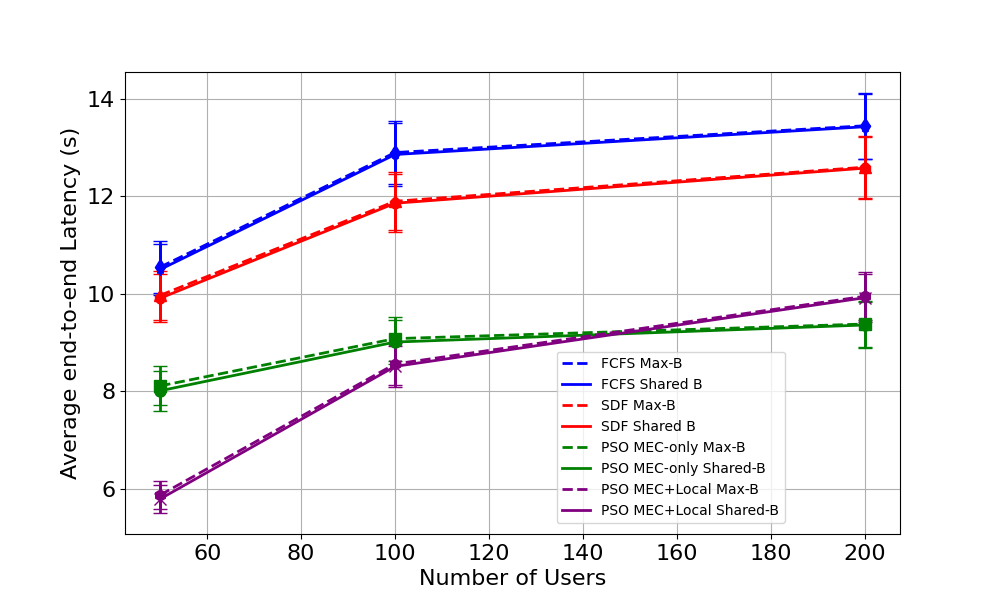}
    \caption{Average E2E latency for different number of users}
    \label{fig:final_L}
\end{figure}

\begin{figure}[!hbt]
    \centering
    \includegraphics[width=0.35\textwidth, clip]{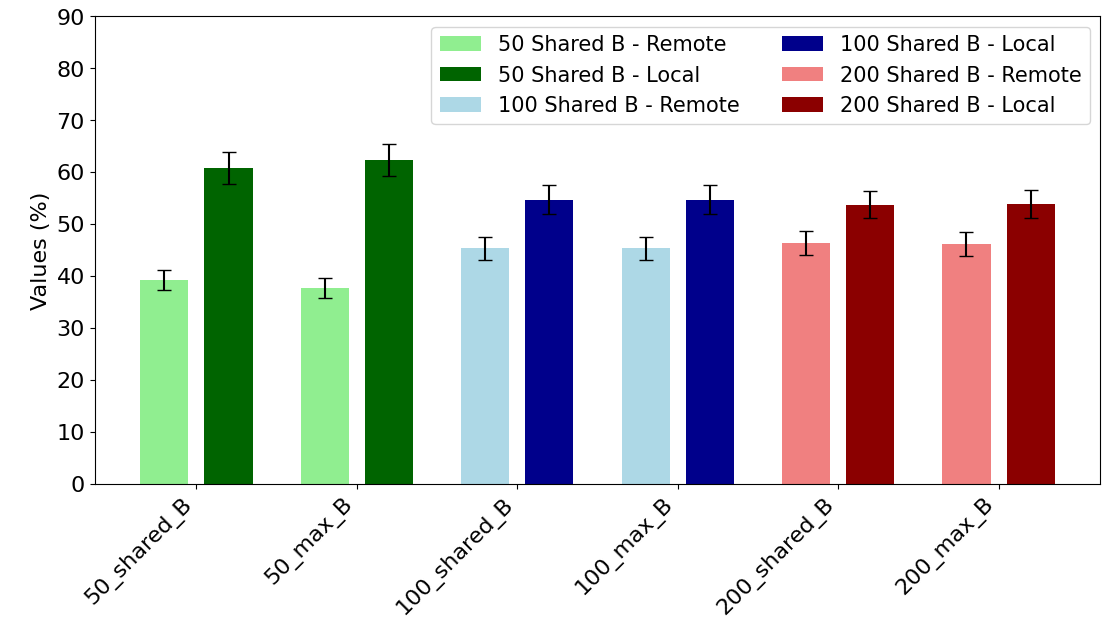}
    \caption{Average local and remote portions for different number of users}
    \label{fig:Local and remore portions}
\end{figure}

The parameters for simulation and PSO are listed in Table~\ref{tab:parameters}. The maximum bandwidth is 20 MHz, but after accounting for a 4.6\% guard band, the effective usable bandwidth is 19.08 MHz.
Figure~\ref{fig:Con_mec_only} and Figure~\ref{fig:Con_partial} present the convergence plots for PSO in the MEC-only and MEC+Local (partitioning) scenarios, illustrated for the maximum number of vehicles and shared bandwidth. The number of iterations in each figure varies due to early stopping. For better comparison, we present the ratio of dropped tasks in Figure~\ref{fig:Drop Task Ratio}, defined as the number of dropped tasks divided by the total number of tasks. The figure demonstrates that as the number of tasks increases, preventing task drops becomes challenging due to short deadlines and high waiting times caused by the larger task load. To address this, we implement MEC+Local (partitioning) approach . As shown, in the MEC+Local (partitioning) approach, no tasks are dropped, indicating that PSO successfully identifies the optimal number of portions to offload, ensuring tasks meet their deadlines while maximizing the offloaded portions as much as possible. \figurename~\ref{fig:final_L} displays the average of end-to-end latency for 10 runs. As shown, with an increasing number of tasks, MEC+Local (partitioning) scenario does not reduce the latency compared to MEC-only offloading. This result arises from the significant difference in numbers of dropped tasks between MEC-only and MEC+Local (partitioning) approaches, with a difference of 109 tasks out of 200. As dropped tasks' latencies are not included in the end-to-end latency, latency is lowered in the MEC-only scenario. In FCFS and SDF, where the order of tasks is handled simply, both the number of dropped tasks and end-to-end latency are less beneficial compared to PSO. In both approaches, PSO outperforms FCFS and SDF due to its optimized task ordering.
As illustrated in Figure~\ref{fig:Local and remore portions}, most portions are processed locally due to the significant waiting times experienced on the MECs. Without incorporating computational waiting times into the objective function, we observe that remote portions exceed local ones, though this would lead to increased end-to-end latency, highlighting the importance of considering waiting times in the objective function.

\section{Conclusion}
In conclusion, this study shows that task optimization and partitioning improve VEC performance, with shared bandwidth among vehicles offering slight benefits over fixed maximum bandwidth. Findings indicate that, for the MEC-only scenario with 200 tasks using the shared bandwidth approach, 109 tasks are dropped, whereas 0 tasks are dropped in the partitioning approach. 
The results demonstrate that in the scenario involving MEC alone with 200 tasks, the shared bandwidth strategy leads to the dropping of 109 tasks. Conversely, the proposed MEC+Local (partitioning) strategy effectively prevents the dropping of any tasks. This highlights the efficiency of the partitioning approach in managing task distribution compared to the MEC-only scenario. This significant number of dropped tasks in the MEC-only scenario results in a lower 
 average end-to-end latency (5.65\% reduction) since only successfully completed tasks are considered. However, despite this, the end-to-end latency in the partitioning approach is still lower compared to the FCFS and SDF methods. In the future work, we plan to incorporate energy optimization and implement more advanced algorithms to enhance performance and improve system efficiency. Energy optimization is essential to reduce power consumption, particularly in resource-constrained environments, while advanced algorithms can further optimize the overall offloading.


\section*{Acknowledgment}
This work was supported in part by funding from the Innovation for Defence Excellence and Security (IDEaS) program from the Department of National Defence (DND), in part by Natural Sciences and Engineering Research Council of Canada (NSERC) CREATE TRAVERSAL Program, and in part by the NSERC DISCOVERY Program.
\bibliographystyle{IEEEtran}

\end{document}